\documentclass[aps,pra,english,twocolumn,showpacs,preprintnumbers,amsmath,amssymb,floatfix,superscriptaddress]{revtex4-1}
\usepackage[T1]{fontenc}
\usepackage{graphicx}
\usepackage{amssymb}
\usepackage{babel}


\usepackage{babel}
\makeatother

\begin{document}

\title{ Casimir forces in a Plasma: Possible Connections  to Yukawa Potentials}

\author{Barry W. Ninham}
\email{Barry.Ninham@anu.edu.au}
\affiliation{Department of Applied Mathematics, Australian National University, Canberra, Australia}

\author{Mathias Bostr{\"o}m}
\email{mathias.bostrom@smn.uio.no}
\affiliation{Centre for Materials Science and Nanotechnology, University of Oslo, P.O. Box 1048 Blindern, NO-0316 Oslo, Norway}
\affiliation{Department of Energy and Process Engineering, Norwegian University of Science and Technology, NO-7491 Trondheim, Norway}

\author{Clas Persson}
\affiliation{Department of Physics, University of Oslo, P. O. Box 1048 Blindern, NO-0316 Oslo, Norway}
\affiliation{ Department of Materials Science and Engineering, Royal Institute of Technology, SE-100 44 Stockholm, Sweden}
\affiliation{Centre for Materials Science and Nanotechnology, University of Oslo, P.O. Box 1048 Blindern, NO-0316 Oslo, Norway}

\author{Iver Brevik}
\email{iver.h.brevik@ntnu.no}
\affiliation{Department of Energy and Process Engineering, Norwegian University of Science and Technology, NO-7491 Trondheim, Norway}

\author{Stefan Y. Buhmann}
\affiliation{Physikalisches Institut, Albert-Ludwigs-University Freiburg, Hermann-Herder-Str. 3, 79104 Freiburg, Germany}

\author{Bo E. Sernelius}
\email{bos@ifm.liu.se}
\affiliation{Division of Theory and Modeling, Department of Physics, Chemistry and Biology, Link\"{o}ping University, SE-581 83 Link\"{o}ping, Sweden}

\begin{abstract}
We present theoretical and numerical results for the screened Casimir effect between perfect metal surfaces in a plasma.
 We  show how the Casimir effect in an electron-positron plasma can provide an important contribution to nuclear interactions. Our results suggest that there is a  connection between Casimir forces and nucleon forces mediated by mesons.  Correct nuclear energies and meson masses appear to emerge naturally from the screened Casimir-Lifshitz effect.
\end{abstract}


\maketitle

\section{Introduction}

The Casimir and Lifshitz theories of intermolecular (dispersion)
forces\,\cite{Casimir,Lond,Dzya} have occupied such a vast literature
that little should remain to be
said.\,\cite{Ninhb,Pars,Milt,Milt2,Sern} However, there exist still
many gaps in our knowledge of the theory of dispersion forces.
For instance, we will show in this paper that
the presence of any non-zero plasma density between
two perfectly reflecting
plates fundamentally alters
their
long-range Casimir interaction. Such finite plasma densities are
always present near metal surfaces. These results are
discussed in detail
in Sec. II where we give theoretical and numerical results for the
Casimir interaction between two perfect metal surfaces in the presence
of a plasma.

The importance of Casimir forces for electron stability
\cite{CasElec,Boyer,Davies,Balian,Milton}, particle physics, and in
nuclear interactions \cite{NinBosNuc}, has been predicted over the
years. The problem we intend to revisit is similar in spirit to the
old story called "the Casimir mousetrap" for the stability of charged
electrons.\,\cite{Boyer,Milton} The negative charges on an electron
surface give rise to a repulsive force between the different parts of
the surface that has to be counteracted by an attractive force in
order for the electron to have a finite radius. Casimir proposed that
such attractive Poincar{\'e} stresses could come from the zero-point
energy of electromagnetic vacuum fluctuations.\,\cite{CasElec} A
number of attempts have been made to compute such Casimir
energies.\,\cite{Boyer,Davies,Balian,Milton} However, all concluded
that while the magnitude of the interaction was correct, it had the
wrong sign. Further it gave a repulsive
force.\,\cite{Boyer,Davies,Balian,Milton}

Finite plasma densities are present between nuclear particles due to
the presence of the plasma of the fluctuating electron-positron pairs
constantly created and annihilated.
The magnitude and asymptotic form of the screened Casimir potential
between
reflecting
surfaces in the presence of this electron-positron plasma suggest a
possible connection between Casimir forces and nucleon forces.
\,\cite{NinBosNuc} In Sec. III we proceed to explore
this
intriguing
similarities of the screened Casimir potential with
the Yukawa potential for nuclear particles
as mediated by mesons. Essentially
 correct nuclear energies, meson masses and meson lifetimes appear
to emerge naturally from the Casimir-Lifshitz theory.
When taken at face value, the screened-Casimir model of the Yukawa
potential would offer an alternative explanation of nuclear forces as being
due to virtual electron-positron excitations.

A somewhat complementary effect is the Casimir force due to electronic
wave-function overlap as discussed in Ref.~\cite{Villa}. In the latter
case, the force results from real plate electrons whose evanescent
wave functions exponentially decay into the gap between the plates. On
the contrary, in our scenario virtual electron--positron pairs in the
space between the plates mediate the force.

\section{Casimir Effect between perfect metal surfaces in the presence of a plasma}

Consider the Casimir-Lifshitz interaction between  ideal metal
surfaces separated by a plasma of dielectric permittivity

\begin{equation}
\varepsilon (i\omega ) = 1 + \frac{{4\pi \rho {e^2}}}{{m{\omega ^2}}}
= 1 + \frac{\omega_p^2}{\omega ^2},
\label{Eq1}
\end{equation}
where the plasma frequency is identified as $\omega_p^2=4 \pi \rho
e^2/m$, $\rho$ is the number density of the plasma, $m$ the electron
mass, and $e$ the unit charge. We define some additional variables
$\bar \rho  = \rho {e^2}{\hbar ^2}/\left( {\pi m{k^2}{T^2}} \right)$,
$\kappa=\omega_p/c$ (note the occurrence of a factor $m c^2$ in the
screening parameter $\kappa$), and $x=2 k T l/(\hbar c)$.
In these expressions  $k$ is Boltzmann's constant, $\hbar$ is
Planck's constant, $T$  the effective temperature of the plasma, $c$ is the
velocity of light, and $l$ the distance between the plates. The exact
expressions for the Casimir-Lifshitz free energy between both real and
perfect metal surfaces across a plasma are given in  Appendix A. We
have found (see  Appendix B for a derivation) that the asymptotic
interaction energy can at high temperatures and/or large separations
be written as

\begin{equation}
F(l,T)=F_{n=0}+F_{n>0},
\label{Eq55}
\end{equation}

\begin{equation}
F_{n=0}(l,T) \approx  - \frac{{kT{\kappa ^2}}}{{2\pi }}{e^{ - 2l\kappa }}[\frac{1}{{2l\kappa }} + \frac{1}{{4{l^2}{\kappa ^2}}}],
\label{Eq56}
\end{equation}

\begin{equation}
{F_{n > 0}} \approx \frac{{{{(kT)}^2}}}{{l\hbar c}}{e^{ - \pi \bar \rho x}}{e^{ - 2\pi x}} + O({e^{ - {x^2}}}).
\label{Eq57}
\end{equation}
Here we have separated the zero and finite frequency contributions.
These expressions may be useful for theoretical comparisons with
experimentally measured Casimir-Lifshitz
forces\,\cite{Lamo,BoSer,Svet,Bord,Klim,Hoye,Schw,Milt,Sushkov,Milt,
Mund} between metal surfaces interacting across a high density plasma.

\begin{figure}
\includegraphics[width=8cm]{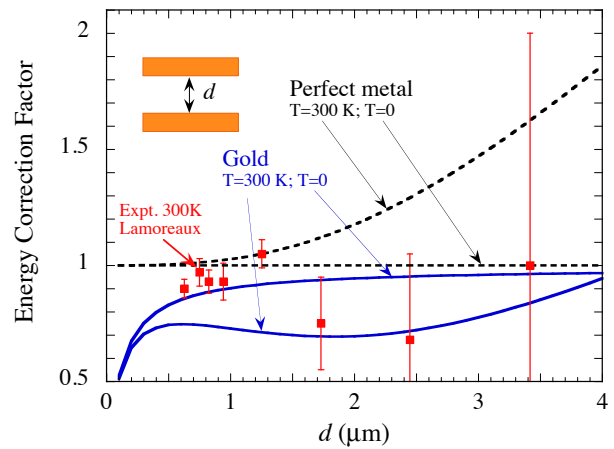}
\caption{Energy correction factor for two gold plates in the absence of any intervening plasma. The filled squares
with error bars are the Lamoreaux' experimental \cite{Lamo} values from the
torsion pendulum experiment.  The dashed curves are the perfect metal
results.  The thick solid curves are the results for real gold plates at zero temperature and at room
temperature\,\cite{BoSer}. The dielectric properties for gold was obtained from tabulated
experimental optical data.}
\label{figu1}
\end{figure}
We first recall  the present understanding of Casimir effect between
real metal surfaces in the absence of any intervening plasma.
Fig.\,\ref{figu1} shows the experimental result of
Lamoreaux\,\cite{Lamo}, compared to the theoretical results of
Bostr{\"o}m and Sernelius\,\cite{BoSer}. All curves show the
interaction energy divided by the result of the ideal Casimir gedanken
experiment at zero temperature, $- \hbar c{\pi ^2}/\left( {720{d^3}}
\right)$. The lowest curve is for gold at room temperature. It was
derived using tabulated optical data for gold as input. Use of the
Drude model gives overlapping results. To be noted is that theory and
experiment clearly disagree for the cluster of experimental points
around $d = 1 {\mu}m$. The experimental results agree better with the
zero-temperature results (upper solid curve) and even with the zero-
or finite temperatures results for ideal metals (the Casimir gedanken
experiment, dotted curves). The agreement is even better with the
theoretical room temperature result obtained when using the plasma
model.

This puzzling behavior has given rise to a long-standing controversy
in the field. We note that the zero frequency part of the Casimir
interaction between real metal surfaces depends on how the dielectric
function of the metal surfaces is treated. Different theoretical
groups have found very different
results.\,\cite{BoSer,Hoye,Svet,Bord,Klim}. A most valuable property
of the Lamoreaux experiment \cite{Lamo} was that it was carried out at
large separations. Lamoreaux was also involved in a more recent
version of his old experiment \cite{Sushkov} (cf. also the comments of
Milton \cite{Milt2}), where plate separations between 0.7 and 7 $\mu$m
were tested. Quite convincingly, the theoretical predictions based
upon the Drude model were found to agree with the observed results to
high accuracy.

The thermal Casimir effect is however a many-facetted phenomenon and
care has to be taken about the electrostatic
patch potentials, which cause uncertainties in the interpretation of
the data in the mentioned experiment. There are other experiments, in
particular the very accurate one of Decca {\it et al.} \cite{Decca},
which yield results apparently in accordance with the plasma model
rather than the Drude model. The reason for this conflict between
experimental results is not known in the community. It has been
suggested occasionally that it might have something to do with the
so-called Debye shielding, which can change the effective gap between
plates from the geometrically measured width. But the experimentalists
themselves turn out to be skeptical towards such a possibility. (An
elementary overview of the temperature dependence of the Casimir force
is recently given in Ref.~\cite{Brevik14}.)  There is clearly an
urgent need for more experiments and theoretical analysis focusing on
Casimir-Lifshitz forces in different systems that include metal
surfaces.

\begin{figure}
\includegraphics[width=8cm]{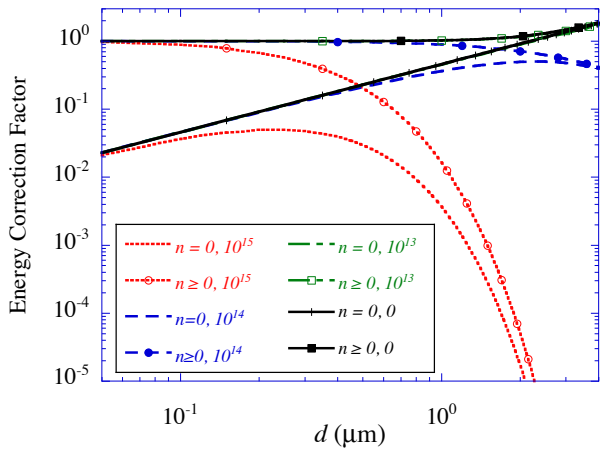}
\caption{Energy correction factor for two perfect metal plates interacting across a plasma.  The curves are the results for perfect metal plates at room temperature for different plasma frequencies ($\omega_p$ in units of rad/s) for the intervening plasma. We show the results for the $n=0$ and $n\geq0$ contributions to the interaction free energy.}
\label{figu2}
\end{figure}

\begin{figure}
\includegraphics[width=8cm]{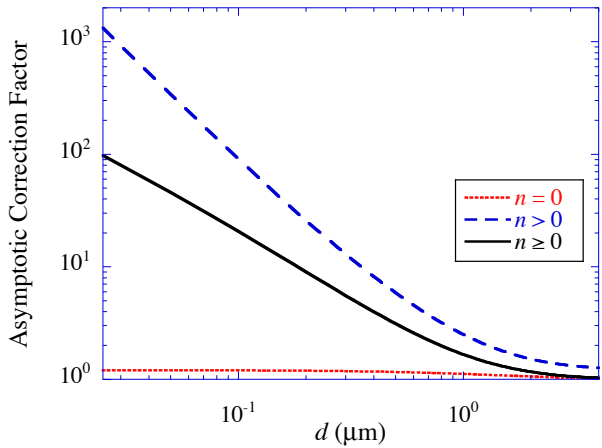}
\caption{Asymptotic  correction factor for two  perfect metal plates interacting across a plasma ($\omega_p=10^{14} rad/s$).  The results show the ratio between numerically calculated energies and the corresponding asymptotes given in the text. There is very good agreement (ratio close to one) for the $n=0$ contribution in the entire range considered. For  $n>0$ and $n\geq0$ the curves  go towards one at large surface separations. Note that the asymptotes become more accurate for higher plasma densities.}
\label{figu3}
\end{figure}

As we have shown so far,  the presence of any intervening
plasma is of importance for the long range interaction energy. We
explore next the effect on the energy correction factor for different
plasma densities between two ideal surfaces, See  Fig.\,\ref{figu2}.
Again, all curves show the interaction energy divided by the result of
the ideal Casimir gedanken experiment at zero temperature in the
absence of a plasma. At large separation the result is strongly
influenced by intervening plasma,
leading to a considerable reduction of the interaction energy. The
results show that even weak intervening plasmas can strongly affect
Casimir force measurements. The possible presence of spurious plasma
densities thus has to be considered carefully.

We next investigate the accuracy of the asymptotes~(\ref{Eq56}) and
(\ref{Eq57}) by comparing their predictions with exact numerical
results.

 Fig.\,\ref{figu3} shows the ratio between numerically
calculated free energy between two perfect metal plates across a
plasma ($\omega_p=10^{14} \mathrm{rad}/s$) to the corresponding
asymptotes given in the text. We see that in this case the asymptote
for the $n=0$ term is very accurate. For the $n>0$ and $n\geq0$
contributions this ratio only goes towards one at large separations.
 The asymptotes become for higher plasma densities.

\section{A contribution from screened Casimir interaction in nuclear interactions}

We will now point out a potential connection with the meson theory.  That
is, if we take the Casimir expansion without  a plasma, the first
three terms (see  Eq. (7) below) are: (1) the usual  zero point fluctuation energy (also
equivalent to current-current correlations); (2) a "chemical
potential" term,  identifiable as the energy of an  electron positron
pair sea (see Landau and Lifshitz\,\cite{Landau}); (3) the black body
radiation in the gap. One can then ask how electromagnetic (EM) theory
can give rise to weak interactions  of particle physics. Such
contribution from EM theory comes out  if one equates the zero point
energy to the black body radiation term. That gives an equivalent
density for the electron positron pair sea and the energy of
interaction of about 8 Mev. This agrees with   the experimentally found
nuclear interaction energy. The form of the  interaction with a plasma
in the gap is the same as that  for the Klein-Gordon--Yukawa potential
with the plasma excitation corresponding to and identical with the
$\pi_0$ meson mass. (This assumes a plate size of 1 Fermi squared in
area and that the planar results translated roughly over to that for
spheres.)

Now we will explore these ideas in more detail. The screened Casimir free energy
asymptotes in the previous section can be compared with the Yukawa
potential between nuclear particles  at distances
large compared with the screening length  $l_\pi=\hbar/m_\pi c$
($m_\pi$  is the mass of the mediating meson),

\begin{equation}
F(l,T) \propto e^{-l/l_\pi}.
\label{Eq58}
\end{equation}

To test if the idea can be correct, we first extract the meson mass by
taking the exponents in the $F_{n=0}$ term given in the previous
section and the Yukawa potential to be equal:

\begin{equation}
{m_\pi } = \frac{{4e\hbar }}{{{c^2}}}\sqrt {{\frac {\pi ({\rho _ + } + {\rho _ - })} {m}}}.
\label{Eq59}
\end{equation}

Since
we know the meson mass (135 MeV) we  estimate
the screening length to be 1.458 fm and we also find the density of
electrons and positrons
that would be required to generate this Yukawa potential from the
Casimir effect.
The equilibrium of electron positron
production can at high temperatures be written as  $\rho_\pm =3
\zeta(3) k^3 T^3/(2 \pi^2 \hbar^3 c^3)$.\,\cite{Landau}
This means that the
required
effective temperature of nuclear interaction via a
screened Casimir interaction is $3.2 \times 10^{11} $ K.

We now address the important question  where the energy to generate
this local electron plasma can come from.\,\cite{NinBosNuc} Feynman
speculated that high energy potentials could excite states
corresponding to other eigenvalues, possibly thereby corresponding to
different masses.\,\cite{Feynman}
It turns out that the low-temperature
Casimir interaction, i.e., without an intervening plasma, by itself
could be capable of generating the effective temperature required to
obtain the plasma. The connection between temperature and density of
electrons and positrons given above is exploited in the expression for
low temperature Casimir interaction between perfectly reflecting
surfaces. In the absence of an intervening plasma, it can be written
as Eq.(35) in Ref. \cite{NinDai}:

\begin{equation}
F(l,T) \approx \frac{{ - {\pi ^2}\hbar c}}{{720{l^3}}} - \frac{{ \zeta (3) k^3 T^3}}{2 \pi \hbar^2 c^2} + \frac{{{\pi ^2}l{k^4}{T^4}}}{{45{\hbar ^3}{c^3}}}.
\label{Eq60a}
\end{equation}
This  can further be re-written as
\begin{equation}
F(l,T) \approx \frac{{ - {\pi ^2}\hbar c}}{{720{l^3}}} - \frac{{\pi ({\rho _ - } + {\rho _ + })\hbar c }}{6} + \frac{{{\pi ^2}l{k^4}{T^4}}}{{45{\hbar ^3}{c^3}}},
\label{Equ60b}
\end{equation}
where the first term is the zero-temperature Casimir energy, the third is the blackbody energy, and the second has been rewritten in terms of electron and positron densities. If we assume that the entire zero-temperature
Casimir energy is transformed into blackbody energy (which at high temperatures can generate an electron-positron plasma) we can estimate the temperature as $T \approx \hbar c/2 l k$. This will at a distance of 3.6 Fermi give the required effective temperature (at the distances discussed above the effective
temperature is even larger, around $2.3\times 10^{12}$ K). It is intriguing
that a cancellation of the Casimir zero point energy and the blackbody energy term, just like the cancellation of the
$n=0$ term at low temperatures, gives the right result.

The screened Casimir interaction between two perfectly reflecting surfaces, with estimated cross section of 1 fermi squared a distance $0.5$ Fermi apart, receives around 4.25 MeV from the n=0 term and 3.25 MeV from the n>0 terms. While the screening length of the n=0 term is defined above we find that the screening length of the n>0 terms also comes out of the right order of magnitude (it is within the crude approximations made of the order one fermi). The nuclear interaction as a screened Casimir interaction would thus receive approximately equal contributions from the classical (n=0) and quantum (n>0) terms. The result compares remarkably well with the binding energy of nuclear interaction that is around 8 MeV.

 If the arguments we have given connecting nuclear and electromagnetic interactions have any substance,  it is hard to avoid the speculation that the standard decomposition in nuclear physics into coulomb and nuclear force contributions may not be entirely correct.
  In the insightful words of Dyson: "The future theory will be built, first of all upon the results of future experiments, and secondly upon an understanding of the interrelations between electrodynamics and mesonic and nucleonic phenomena".\,\cite{Dyson} The problem is precisely equivalent to that which occurs in physical chemistry where standard theories have all been based on the ansatz that electrostatic forces (treated in a nonlinear theory) and electrodynamic forces (treated in linear approximation by Lifshitz theory) are separable. The ansatz violates both the Gibbs adsorption equation and the gauge condition on the electromagnetic field.\,\cite{NinYam} When the defects are remedied a great deal of confusion appears to fall into place.  If these results are not acceptable within the standard model one must still consider the presence of this additional electromagnetic fluctutation interaction energy between nuclear particles.

\section{Conclusions}

We have explored the effect of an intervening plasma on the Casimir
force between two perfectly conducting plates. The analytically
derived asymptotes for large plate separations show that even spurious
plasma densities can considerably reduce the expected Casimir force.
The effect of plasmas should therefore carefully be considered in
Casimir-force measurements.

In addition, the derived asymptotes show an interesting structural
analogy with the Yukawa potential of nuclear interactions. We have
explored this analogy to discuss whether the electromagnetic Casimir
effect can possibly explain these interactions. The comparison yields
predictions for the required virtual electron-positron plasma density
which, however, is only achievable at very large ambient temperatures. If the potential connection to nuclear interactions is  correct, then we speculate that the charged $\pi_+$ and $\pi_-$ mesons would come  out to be bound positron-plasmon and electron-plasmon excitations in the  electron-positron plasma.

Apart from these speculations,  our main idea has been to investigate to what extent the screened Casimir effect between perfect metal surfaces, intervened by an
electron-positron plasma, can be applied to estimate nucleon forces
mediated by mesons. Figures 2 and 3 show the effect of plasma
screening; especially the large suppression of the Casimir energy
when the plasma density is large, is clearly shown in Fig.~2. Our
main findings are that nuclear energies and meson masses emerge
numerically of the right order of magnitude, thus indicating that our
basic idea is a viable one.  

   Of course, the ideas explored in this paper are somewhat speculative. In principle, although the Casimir energy has the right order of magnitude to provide the required temperature, one may object that it is not evident how this energy can be converted to thermal radiation. The point we wish to emphasize here is  that the present arguments, although incomplete, may serve as a useful starting point for further research in this direction, perhaps within the framework of quantum statistical mechanics.

A final comment: Use of the electrodynamic Casimir effect in the
context of nuclear physics is of course not new. For instance, in
hadron spectroscopy viewed from the standpoint of the MIT quark bag
model it has long been known that the zero-point fluctuations of the
quark and gluon fields may generate a finite zero-point energy of the
form $E_{\rm zp}=-Z_0/r$, for massless quarks. The constant $Z_0$ is
not firmly grounded theoretically, but serves a a phenomenological
term fitting the experimental data (a classic review article in this
field is that of Hasenfratz and Kuti \cite{Hasenfratz}). The
phenomenological quark model in which the $r-$dependent part $\Delta
m(r)$ of the effective quark mass $m(r)$ varies according to a
Gaussian, $\Delta m(r) \propto -e^{-r^2/R_0^2}$, can also be regarded
as an example of essentially the same kind \cite{Brevik86}.

\appendix
\section{Casimir-Lifshitz Free Energy}

One way to find retarded van der Waals or Casimir-Lifshitz
interactions between two objects interacting across a medium is in
terms of the electromagnetic normal modes of the
system.\,\cite{Lifshitz,Mahanty} For planar structures the
interaction energy per unit area can be written as
\begin{equation}
    E = \hbar \int {\frac{{{d^2}q}}{{{{\left( {2\pi } \right)}^2}}}} \int\limits_0^\infty  {\frac{{d\omega }}{{2\pi }}}     \ln \left[ {{f_q}\left( {i\omega } \right)} \right],
    \label{eq:hamIntEnergyIntegral}
\end{equation}
where $f_q$ is the mode condition function with
${f_q}\left( {{\omega_q}} \right) = 0$ defining electromagnetic normal
modes. Eq.~(\ref{eq:hamIntEnergyIntegral}) is valid for zero
temperature and the interaction energy is the internal energy. At
finite temperature the interaction energy is a free energy and can be
written as
\begin{equation}
    F =
\sum\limits_{n = 0}^\infty F_n=
\frac{1}{\beta}\int {\frac{{{d^2}q}}{{{{\left( {2\pi } \right)}^2}}}}
\sum\limits_{n = 0}^\infty{\!'} {\ln \left[ {{f_q}\left( {i{\xi _n}}
\right)} \right];}
    \label{eq:hamIntEnergySum}
\end{equation}
where $\beta  = 1/{k} T$, and the prime on the summation sign
indicates that the term for $n = 0$ should be divided by two. The
integral over frequency in Eq.~(\ref{eq:hamIntEnergyIntegral})
has been replaced by a summation over discrete Matsubara frequencies
\begin{equation}
    \xi_n = \frac{2\pi n}{\hbar \beta}; \; n = 0,\,1,\,2,\, \ldots
\end{equation}

For planar structures the quantum number that characterizes the normal
modes is {\bf q}, the two-dimensional (2D) wave vector in the plane of
the interfaces. Two mode types can occur: transverse magnetic (TM) and
transverse electric (TE). These dictate the form through the wave
amplitude reflection coefficients, $r$. For instance, for two planar
objects in a medium, corresponding to the geometry 1|2|1, the mode
condition function is given by
\begin{equation}
    f_q = 1 - e^{ - 2 \gamma _2 d} {r_{12}}^2,
\label{eq:modeCondFunct}
\end{equation}
where $d$ is the thickness of intermediate medium, and the reflection
coefficients for a wave impinging on an interface between medium
$1$
and
$2$
from the
$1$-side given as
\begin{equation}
r_{12}^{\mathrm{TM}}
= \frac{{{\varepsilon _2}{\gamma _1} -
{\varepsilon _1}{\gamma _2}}}{{{\varepsilon _2}{\gamma _1} +
{\varepsilon _1}{\gamma _2}}},
    \label{eq:radialTM}
\end{equation}
and
\begin{equation}
r_{12}^{\mathrm{TE}} = \frac{{\left( {{\gamma _1} - {\gamma _2}}
\right)}}{{\left( {{\gamma _1} + {\gamma _2}} \right)}},
    \label{eq:radialTE}
\end{equation}
for TM and TE modes, respectively. Here,
$\gamma_j$
stands for
\begin{equation}
    {\gamma _i} (\omega) = \sqrt {q^2 - {\varepsilon _i}\left( \omega  \right){{\left( {\omega /c} \right)}^2}}. \label{eq:gamma}
\end{equation}
where ${{\varepsilon _i}\left(\omega\right)}$ is the dielectric
function of medium $i$, and $c$ the speed of light in
vacuum.

For two perfectly conducting plates (\mbox{$r_{12}^{\mathrm{TE}}=-1$},
\mbox{$r_{12}^{\mathrm{TM}}=1$}), the Casimir energy
(\ref{eq:hamIntEnergySum}) across a dissipation-free plasma takes
the simple form
\begin{equation}
F(l,T) = \frac{{kT}}{\pi }\sum\limits_{n = 0}^\infty  {'\int_0^\infty
d } qq\ln \left[ {1 - {e^{ - 2l\sqrt {{q^2} + {{({\xi _n}/c)}^2} +
{\kappa ^2}}
}}} \right],\label{EqA11}
\end{equation}
recall that $\kappa=\omega_p/c$.
By a simple variable substitution, the first term in the Matsubara
sum can be cast int the alternative form
\begin{equation}
F_{n=0}(l,T) =   \frac{{kT}}{{2\pi}}  \int_\kappa^\infty dt t
\ln(1-e^{-2 l t}).
\label{EqA10}
\end{equation}

\section{Asymptotic Casimir Free Energy in a Plasma}
Exact treatments of Casimir forces between perfect metal surfaces
across a plasma using the above expressions typically  give asymptotic
expansions that are  not uniformly valid. The treatment we present
here gives a  different result. Our starting point is
the above formula~(\ref{EqA11}) for the ineraction of two perfectly
conducting plates accross an intervening plasma
as re-written by Ninham and Daicic.\,\cite{Mahanty,NinDai,Elizalde}

\begin{equation}
F(l,T) =  - \frac{{ - kT}}{{4\pi {l^2}}}\frac{1}{{2\pi i}}\int_c d s\frac{{\Gamma (s)\zeta (s + 1)}}{{(s - 2){{(2\pi x)}^{s - 2}}}}  \sum\limits_{n = 0}^\infty  {'{{({n^2} + \bar \rho )}^{1 - s/2}}}
\label{Eq2a}
\end{equation}

The zero frequency ($n=0$) term gives the following contribution,
\begin{equation}
F_{n=0}(l,T) =  - \frac{{ - kT}}{{8\pi {l^2}}} {\frac {1} {2 \pi i}} \int_c ds {\frac {\Gamma (s) \zeta (s+1)} {(s-2) (2 \kappa l)^{s-2}}},
\label{Eq2b}
\end{equation}

\begin{equation}
F_{n=0}(l,T) =   \frac{{  kT}}{{2\pi}}  \int_\kappa^\infty dt t
\ln(1-e^{-2 l t}),
\label{Eq2c}
\end{equation}
which at large separations becomes:\,\cite{NinDai}
\begin{equation}
{F_{n = 0}} \approx  - \frac{{kT{\kappa ^2}}}{{2\pi }}{e^{ - 2l\kappa }}[\frac{1}{{2l\kappa }} + \frac{1}{{4{l^2}{\kappa ^2}}}].
\label{Eq2d}
\end{equation}

The interaction free energy can be written as:\,\cite{NinDai}
\begin{equation}
\begin{array}{l}
F(l,T) =  - \frac{{ - kT}}{{8\pi {l^2}}}\frac{1}{{2\pi i}}\int_c d s\frac{{\Gamma (s)\zeta (s + 1)}}{{(s - 2){{(2\pi x)}^{s - 2}}}}\\
\quad \quad \quad \quad \quad \quad \quad \quad \quad  \times {\zeta _G}( - 1 + s/2,\bar \rho ),\;c > 3,
\end{array}
\label{Eq2e}
\end{equation}

\begin{equation}
\zeta_G(z,a)=2 \zeta_{EH} (z,a)+a^{-z}=2 \sum_{n=1}^\infty {\frac {1} {(n^2+a)^z}}+a^{-z}
\label{Eq2f}
\end{equation}
where as discussed in detail by Ninham and Daicic the generalized Epstein-Hurwitz $\zeta$ function $\zeta_G$ is meromorphic
and has simple poles in the complex plane at z=-k+1/2 (k=0,1,2,..) \,\cite{NinDai}.
In the limit of low temperatures or distances $x<<1$ they found  (see Ref. \cite{NinDai}  for the complete expression)
\begin{equation}
F(l,T)={\frac {- \pi \hbar c} {720 l^3}} \left[1-15 {\frac {\rho {e^2}{\hbar ^2}} {\left( {\pi m{k^2}{T^2}} \right)}} ({\frac{2 k T l} {\hbar c}})^2- ...... \right]
\label{Eq2g}
\end{equation}
where at low temperatures the $n=0$ term cancels out a contribution from the higher frequency terms.

It is possible  with some algebra to express the Lifshitz free energy between two  ideal metal plates with intervening plasma in the following form (useful for deriving the asymptotes considered in this contribution):

\begin{equation}
F(l,T) =  - \frac{{ - kT}}{{4\pi {l^2}}}\eta (l,T) = \frac{{ - kT}}{{4\pi {l^2}}}\left[ {\pi {x^3}I\left( {\bar \rho ,x} \right)} \right].
\label{Eq2h}
\end{equation}
The integral $I$ consists of two parts,
\begin{equation}
I= I_1+I_2,
\label{Eq3}
\end{equation}
where
\begin{equation}
I_1= \int_0^\infty  dy e^{- \pi {\bar \rho}  y} y^{-5/2}  {\bar \omega}(x^2/y),
\label{Eq4}
\end{equation}
and
\begin{equation}
{I_2} = \int_0^\infty  d y{e^{ - \pi \bar \rho y}}{y^{ - 5/2}}\bar \omega ({x^2}/y)2\bar \omega (y),
\label{Eq5}
\end{equation}
respectively.

The function $\bar \omega (y)$ appearing in both integrands is defined as,\,\cite{Elizalde}
\begin{equation}
\begin{array}{*{20}{l}}
{\bar \omega (y) \equiv \sum\limits_{n = 1}^\infty  {{e^{ - {n^2}\pi y}}}  \equiv \frac{1}{2}\left\{ { - 1 + {y^{ - 1/2}}\left[ {1 + 2\bar \omega (1/y)} \right]} \right\}.}
\end{array}
\label{Eq6}
\end{equation}
The sum converges faster the larger the $y$-value. To make use of this fact we divide the integration range for $I_1$ into two parts and use the two different expressions for the sum in the two resulting integrals. Thus,
\begin{equation}
I_1= H_1+H_2,
\label{Eq7}
\end{equation}
where
\begin{equation}
{H_1} = \int_{{x^2}}^\infty  d y{e^{ - \pi \bar \rho y}}{y^{ - 5/2}}[\frac{1}{2}( - 1 + \sqrt {y/{x^2}} ) + \sqrt {y/{x^2}} \bar \omega (y/{x^2})],
\label{Eq8}
\end{equation}
and
\begin{equation}
H_2= \int_0^{x^2}  dy e^{- \pi {\bar \rho}  y} y^{-5/2}  {\bar \omega}(x^2/y),
\label{Eq9}
\end{equation}
respectively.

The integrand in $I_2$ has a product of two $\bar \omega$ functions with different arguments. Here we divide the integration range into three regions and choose the form of the sum that gives the fastest convergence. Under the assumption that $x>1$ we have
\begin{equation}
\begin{array}{*{20}{l}}
{{I_2} = \int_0^1 d y{e^{ - \pi \bar \rho y}}{y^{ - 5/2}}\bar \omega ({x^2}/y)[ - 1 + {y^{ - 1/2}}(1 + 2\bar \omega (1/y))]}\\
\begin{array}{l}
 + 2\int_1^{{x^2}} d y{e^{ - \pi \bar \rho y}}{y^{ - 5/2}}\bar \omega ({x^2}/y)\bar \omega (y)\\
 + \int_{{x^2}}^\infty  d y{e^{ - \pi \bar \rho y}}{y^{ - 5/2}}\bar \omega (y)[ - 1 + \sqrt {y/{x^2}} (1 + 2\bar \omega (y/{x^2})]
\end{array}\\
{ = {J_1} + {J_2} + {J_3} + {J_4} + {J_5} + {J_6} + {J_7}},
\end{array}
\label{Eq10}
\end{equation}
where
\begin{equation}
 J_ 1=
2 \int_1^{x^2}  dy e^{- \pi {\bar \rho}  y} y^{-5/2}  {\bar \omega}(x^2/y)  {\bar \omega} (y),
\label{Eq11}
\end{equation}
\begin{equation}
 {J_2} = \frac{2}{x}\int_{{x^2}}^\infty  d y{e^{ - \pi \bar \rho y}}{y^{ - 2}}\bar \omega (y)\bar \omega (y/{x^2}),
\label{Eq12}
\end{equation}
\begin{equation}
 J_ 3=2  \int_0^1 dy y^{-3} {\bar \omega} (1/y)  {\bar \omega} (x^2/y)  e^{- \pi {\bar \rho}  y},
\label{Eq13}
\end{equation}
\begin{equation}
J_4=- \int_{x^2}^\infty dy e^{- \pi {\bar \rho}  y} y^{-5/2}  {\bar \omega} (y),
\label{Eq14}
\end{equation}
\begin{equation}
{J_5} = \frac{1}{x}\int_{{x^2}}^\infty  d y{e^{ - \pi \bar \rho y}}{y^{ - 2}}\bar \omega (y),
\label{Eq15}
\end{equation}
\begin{equation}
J_6=-\int_0^1 dy e^{- \pi {\bar \rho}  y} y^{-5/2}  {\bar \omega}(x^2/y),
\label{Eq16}
\end{equation}
ans
\begin{equation}
J_7=\int_0^1 dy e^{- \pi {\bar \rho}  y} y^{-3}  {\bar \omega}(x^2/y),
\label{Eq17}
\end{equation}
respectively.

We now add the two $I$ terms and recombine the integral terms to find
\begin{equation}
I= I_1+I_2=J_1+J_2+J_3+K_1+K_2+K_3+K_4+K_5,
\label{Eq18}
\end{equation}
where
\begin{equation}
{K_1} = \frac{1}{x}\int_{{x^2}}^\infty  d y{e^{ - \pi \bar \rho y}}{y^{ - 2}}[\frac{1}{2} + \bar \omega (y)],
\label{Eq19}
\end{equation}
\begin{equation}
{K_2} = \frac{{ - 1}}{x}\int_{{x^2}}^\infty  d y{e^{ - \pi \bar \rho y}}x{y^{ - 5/2}}[\frac{1}{2} + \bar \omega (y)],
\label{Eq20}
\end{equation}
\begin{equation}
K_3=J_7,
\label{Eq21}
\end{equation}
\begin{equation}
{K_4} = \frac{1}{x}\int_{{x^2}}^\infty  d y{e^{ - \pi \bar \rho y}}{y^{ - 2}}\bar \omega (y/{x^2}),
\label{Eq22}
\end{equation}
and
\begin{equation}
\begin{array}{l}
{K_5} = \int_1^{{x^2}} d y{e^{ - \pi \bar \rho y}}{y^{ - 5/2}}\bar \omega ({x^2}/y)\\
 = \frac{1}{{{x^3}}}\int_1^{{x^2}} d y{e^{ - \pi \bar \rho {x^2}/y}}\sqrt y \bar \omega (y),
\end{array}
\label{Eq23}
\end{equation}
respectively.

In $K_4$ we let $y \to {x^2}\xi $,

\begin{equation}
{K_4} = \frac{1}{{{x^3}}}\int_1^\infty  d \xi {e^{ - \pi \bar \rho {x^2}\xi }}{\xi ^{ - 2}}\bar \omega (\xi),
\label{Eq24}
\end{equation}
and let $\xi  \to 1/y$, use the definition of ${\bar \omega} (y)$, and separate into three terms,
\begin{equation}
K_4=M_1+M_2+M_3
\label{Eq25}
\end{equation}
where
\begin{equation}
{M_1} = \frac{{ - 1}}{{2{x^3}}}\int_0^1 {dy{e^{ - \pi \bar \rho {x^2}/y}}}  = \frac{{ - 1}}{{2x}}\int_{{x^2}}^\infty  {dy{e^{ - \pi \bar \rho y}}{y^{ - 2}}} ,
\label{Eq26}
\end{equation}
\begin{equation}
{M_2} = \frac{1}{{2{x^3}}}\int_0^1 {dy\sqrt y {e^{ - \pi \bar \rho {x^2}/y}}}  = \frac{1}{2}\int_{{x^2}}^\infty  {dy{e^{ - \pi \bar \rho y}}{y^{ - 5/2}}},
\label{Eq27}
\end{equation}
and
\begin{equation}
{M_3} = \frac{1}{{{x^3}}}\int_0^1 {dy\sqrt y {e^{ - \pi \bar \rho {x^2}/y}}} \bar \omega \left( y \right),
\label{Eq28}
\end{equation}
respectively.

$M_1$ and $M_2$ exactly cancel with the terms with $1/2$ in $K_1$ and $K_2$. Now we combine the expressions for $K_4$ and $K_5$ and insert into the expression for $I$,
\begin{equation}
I= I_1+I_2=J_1+J_2+J_3+K_3+N_1+N_2,
\label{Eq29}
\end{equation}
where
\begin{equation}
{N_1} = \frac{1}{x}\int_{{x^2}}^\infty  d y{e^{ - \pi \bar \rho y}}({y^{ - 2}} - x{y^{ - 5/2}})\bar \omega (y),
\label{Eq30}
\end{equation}
and
\begin{equation}
{N_2} = \frac{1}{{{x^3}}}\int_0^{{x^2}} d y{e^{ - \pi \bar \rho {x^2}/y}}\sqrt y \bar \omega (y),
\label{Eq31}
\end{equation}
respectively.

Of these terms  $J_2$, $J_3$, $K_3$ and the term $\int_{x^2}^\infty$ are all $O(e^{-x^2})$ and we may drop them. So we have apart from a term $O(e^{-x^2})$ the following expressions:

\begin{equation}
 \eta (l,T)=\eta_1+\eta_2=\pi x^3 (I_1+I_2),
\label{Eq32}
\end{equation}
where
\begin{equation}
 \eta_1 \approx\pi \int_0^{x^2} dy \sqrt{y} e^{- \pi {\bar \rho}  x^2/y}  {\bar \omega} (y),
\label{Eq33}
\end{equation}
and since

\begin{equation}
 \eta_1 \approx \pi \int_0^{\infty} dy \sqrt{y} e^{- \pi {\bar \rho}  x^2/y}  {\bar \omega} (y)-O(e^{-x^2}),
\label{Eq34}
\end{equation}
we may write
\begin{equation}
 \eta_1\approx \pi \int_0^{\infty} dy \sqrt{y} e^{- \pi {\bar \rho}  x^2/y}  {\bar \omega} (y).
\label{Eq35}
\end{equation}

Using the definition of ${\bar \omega} (y)$ and the following representation:

\begin{equation}
{e^{ - y}} = \frac{1}{{2\pi i}}\int_{C - i\infty }^{C + i\infty } d p{y^{ - p}}\Gamma (p),{\mathop{\rm Re}\nolimits} (p) = C > 0,
\label{Eq36}
\end{equation}

we obtain

\begin{equation}
{\eta _1} \approx \pi \int_0^\infty  d y\sqrt y \int_{C - i\infty }^{C + i\infty } d p\sum\limits_{n = 1}^\infty  {\frac{{\Gamma (p)}}{{{{({n^2}\pi y)}^p}}}} {e^{ - \pi \bar \rho {x^2}/y}}.
\label{Eq37}
\end{equation}

With a variable substitution $\kappa^2l^2=\pi^2 {\bar \rho} x^2=4 \pi \rho e^2 l^2/(mc^2)$ we find

\begin{equation}
{\eta _1} \approx \frac{1}{{\sqrt \pi  }}\int_0^\infty  d y\sqrt y \int_{C - i\infty }^{C + i\infty } d p\sum\limits_{n = 1}^\infty  {\frac{{\Gamma (p)}}{{{{({n^2}y)}^p}}}} {e^{ - {\kappa ^2}{l^2}/y}}
\label{Eq38}
\end{equation}
and using the Riemann $\zeta$ function,
\begin{equation}
{\eta _1} \approx \frac{1}{{\sqrt \pi  }}\int_{C - i\infty }^{C + i\infty } d p\Gamma (p)\zeta (2p)\frac{{{{(\kappa l)}^3}}}{{{{(\kappa l)}^{2p}}}}\int_0^\infty  d y{y^{p - 5/2}}{e^{ - y}}.
\label{Eq39}
\end{equation}

Integration over $y$ results in

\begin{equation}
{\eta _1} \approx \frac{{{{(\kappa l)}^3}}}{{\sqrt \pi  }}\int_{C - i\infty }^{C + i\infty } d p\frac{{\Gamma (p)\zeta (2p)}}{{{{(\kappa l)}^{2p}}}}\Gamma (p - \frac{3}{2}),{\rm{Re}}(p) = C > \frac{3}{2}.
\label{Eq40}
\end{equation}

We now exploit relations for the $\Gamma$ function:

\begin{equation}
\Gamma (p - 3/2) = \frac{{\Gamma (p + 1/2)}}{{(p - 1/2)(p - 3/2)}},
\label{Eq41}
\end{equation}

and
\begin{equation}
  \Gamma (p) \Gamma(p+1/2)=\sqrt{\pi} 2^{1-2p} \Gamma(2 p),
\label{Eq42}
\end{equation}
to obtain
\begin{eqnarray}{}
{\eta _1} &\approx &4{(\kappa l)^2}\int_{c - i\infty }^{c + i\infty } d p\frac{{\Gamma (p)\zeta (p + 1)}}{{{{(2\kappa l)}^p}(p - 2)}}\nonumber\\
 &= &4{(\kappa l)^2}\int_{c - i\infty }^{c + i\infty } d p\Gamma (p)\sum\limits_{n = 1}^\infty  {\frac{1}{{{n^{p + 1}}{{(2\kappa l)}^p}(p - 2)}}}\nonumber \\
 &=& 2{(\kappa l)^2}\int_{c - i\infty }^{c + i\infty } d p\Gamma (p)\int_0^\infty  {\frac{{dxx}}{{{{({x^2} + 1)}^{p/2}}{{(2\kappa l)}^p}}}}\nonumber \nonumber \\
 &&\times \sum\limits_{n = 1}^\infty  {\frac{1}{{{n^{p + 1}}}}\nonumber}\\
 &=& 2{(\kappa l)^2}\int_0^\infty  d xx\sum\limits_{n = 1}^\infty  {{n^{ - 1}}} {e^{ - 2\kappa ln\sqrt {{x^2} + 1} }\nonumber}\\
 &=&  - 2{(\kappa l)^2}\int_0^\infty  d xx\ln (1 - {e^{ - 2\kappa l\sqrt {{x^2} + 1} }})\nonumber\\
 &=&  - 2{l^2}\int_\kappa ^\infty  d tt\ln (1 - {e^{ - 2lt}}).
\end{eqnarray}

The free energy from $\eta_1$ is then seen to give a contribution equal to the zero frequency part of the Lifshitz-Casimir energy between  ideal metal surfaces with an intervening plasma,\,\cite{NinDai}

\begin{equation}
{F_1}(l,T) = \frac{{kT}}{{2\pi }}\int_\kappa ^\infty  d tt\ln(1 - {e^{ - 2lt}}).
\label{Eq44}
\end{equation}

The remaining $\eta$ term is

\begin{equation}
 \eta_2 \approx 2 \pi x^3 \int_1^{x^2} dy y^{-5/2} e^{- \pi {\bar \rho}  y}  {\bar \omega} (y) {\bar \omega} (x^2/y).
\label{Eq45}
\end{equation}

Now, since

\begin{equation}
{\bar \omega} (y)=\sum_{n=1}^\infty  e^{-n^2 \pi y}=\sum_{n=0}^\infty  e^{-\pi (n+1)^2 y},
\label{Eq46}
\end{equation}

we have

\begin{equation}
{e^{ - \pi y}} < \bar \omega (y) < \frac{{{e^{ - \pi y}}}}{{1 - {e^{ - 2\pi y}}}},
\label{Eq47}
\end{equation}
and

\begin{equation}
\begin{array}{l}
\int_1^{{x^2}} d y{y^{ - 5/2}}{e^{ - \pi \bar \rho y}}{e^{ - \pi y}}{e^{ - \pi {x^2}/y}}\\
 < {I_2}\\
 < \int_1^{{x^2}} d y{y^{ - 5/2}}\frac{{{e^{ - \pi \bar \rho y}}{e^{ - \pi y}}{e^{ - \pi {x^2}/y}}}}{{(1 - {e^{ - 2\pi y}})(1 - {e^{ - 2\pi {x^2}/y}})}}\\
 < \int_1^{{x^2}} d y{y^{ - 5/2}}\frac{{{e^{ - \pi \bar \rho y}}{e^{ - \pi y}}{e^{ - \pi {x^2}/y}}}}{{{{(1 - {e^{ - 2\pi }})}^2}}}.
\end{array}
\label{Eq48}
\end{equation}

Apart from a very small uncertainty $ (1-e^{- 2 \pi})^2$ we have

\begin{equation}
\eta_2 \approx 2 \pi x^3  \int_1^{x^2} dy y^{-5/2} e^{- \pi {\bar \rho}  y} e^{- \pi  y} e^{- \pi x^2/y},
\label{Eq49}
\end{equation}
and with the substitution $y \to yx$ we have
\begin{equation}
\approx 2 \pi x^{3/2}  \int_{1/x}^{x} dy y^{-5/2} e^{- \pi {\bar \rho}  x y} e^{- \pi  (y+1/y) x},
\label{Eq50}
\end{equation}
which for $x \rightarrow \infty$ (large separations or high temperatures) produces a simple final expression. To find this we notice that the integral has a steep maximum. Take $f(y)=y+1/y$, then $f'(y)=1-1/y^2$ is equal to zero at $y_0=1$ and $f(y_0)=2$ and $f''(y_0)=2$. Thus, we may write

\begin{equation}
\eta_2 \approx 2 \pi x^{3/2}  \int_{-\infty}^{\infty} dy e^{-\pi {\bar \rho}  x } e^{-2 \pi x} e^{- \pi x (y-y_0)^2},
\label{Eq51}
\end{equation}
and
\begin{equation}
\eta_2 \approx 4 \pi x^{3/2}  e^{-\pi {\bar \rho}  x } e^{-2 \pi x}  \int_{0}^{\infty} dt e^{- \pi x t^2}=2 \pi x   e^{-\pi {\bar \rho}  x } e^{-2 \pi x}.
\label{Eq52}
\end{equation}

The free energy from $\eta_2$  gives a contribution at high $x$(large separations or high temperatures)

\begin{equation}
\begin{array}{l}
{F_2} = \frac{{ - {{(kT)}^2}}}{{l\hbar c}}{e^{ - \bar \rho x}}{e^{ - 2\pi x}}\\
\quad  = \frac{{ - {{(kT)}^2}}}{{l\hbar c}}{e^{ - 2\rho \hbar c\frac{{{e^2}}}{{m{c^2}}}l}}{e^{ - 4\pi kTl/(\hbar c)}}.
\end{array}
\label{Eq53}
\end{equation}

The whole Casimir free energy in the high $x=2 k T l/(\hbar c)$ limit is

\begin{equation}
\begin{array}{l}
F(l,T) = \frac{{kT}}{{2\pi }}\int_\kappa ^\infty  d tt\ln(1 - {e^{ - 2lt}})\\
 \quad - \frac{{{{(kT)}^2}}}{{l\hbar c}}{e^{ - 2\rho \hbar c\frac{{{e^2}}}{{m{c^2}}}l}}{e^{ - 4\pi kTl/(\hbar c)}} + O({e^{ - {x^2}}}).
\end{array}
\label{Eq54}
\end{equation}

This is the correct limit for either high temperature at fixed separation or for large distances at fixed temperature.
The given expression can also be valid at small separations or low temperatures. This is a crusial point but one should remember that the derivation of plasma density from the equating of black body radiation to zero point energy and subsequent use of that density requires "high" temperatures.\,\cite{Landau} The situation for two nuclear particles is one with very high effective temperature and separations being "large", at least compared to the screening length of the high density plasma.

\section{$\zeta$ functions in physics}
 We would like to point out that that zeta functions have been applied to many physical problems in the past.\,\cite{Elizalde,ElizaldePRD,Weldon,Actor,Blau} Elizalde considered for example the sum $S_2(t)$, defined by
\[ S_2(t)=\sum_{n=1}^\infty e^{-n^2t}, \]
with $t$ a parameter. This is transformed into the equation
\[ S_2(t)=-\frac{1}{2}+\frac{1}{2}\sqrt{\frac{\pi}{t}}+\sum_{k=1}^\infty \frac{(-t)^k}{k!}\zeta(-2k)+\Delta_2(t), \]
where $\Delta_2(t)$ is a remainder. The zeta-function term does not contribute, and the reminder reduces to the sum/integral
\[ \Delta_2(t)= 2\sum_{n=1}^\infty \int_0^\infty dx e^{-x^2t}\cos(2\pi nx)=\sqrt{\frac{\pi}{t}}\sum_{n=1}^\infty e^{-\frac{\pi^2n^2}{t}}. \]
It means that
\[S_2(t)= -\frac{1}{2}+\frac{1}{2}\sqrt{\frac{\pi}{t}} +\sqrt{\frac{\pi}{t}}\sum_{n=1}^\infty e^{-\frac{\pi^2n^2}{t}}. \]
This formula  was a key component in our derivations.\,\cite{Whittaker}

\acknowledgments{MB and CP acknowledge support from  the Research Council of Norway
(Contract No. 221469). MB also thanks the Department of Energy and
Process Engineering (NTNU, Norway) for financial support. CP thanks the Swedish Research
Council (Contract No. C0485101) for financial support.
This work was supported by the DFG (grant BU 1803/3-1). We thank Dr John Lekner for pointing out the relevance for the analysis of  the Poisson-Jacobi formula (page 124, example 18 in Whittaker and Watson) .\,\cite{Whittaker}}

\end{document}